\documentclass[11pt,a4paper]{article}
\usepackage{jinstpub}
\usepackage[export]{adjustbox}
\usepackage{makecell}
\usepackage{caption}
\usepackage{subcaption}

\usepackage[linesnumbered,boxed]{algorithm2e}
\DontPrintSemicolon

\title{TCT investigation of the one-sided depletion of low-temperature covalently bonded silicon sensor P-N diodes}

\author[1]{J.~W\"uthrich%
\note{Corresponding author.},}
\author{C. Alt}
\author{and A.~Rubbia}

\affiliation{Institute for Particle Physics and Astrophysics, ETH Z\"urich,\\
Otto-Stern-Weg 5, Z\"urich, Switzerland}

\emailAdd{jwuethri@phys.ethz.ch}

\abstract{
In the context of particle detectors, low-temperature covalent wafer-wafer bonding allows for integration of high-Z materials as absorbing layers with readout chips produced in standard CMOS processes.
This enables for instance the fabrication of novel highly efficient X-ray imaging sensors.
In order to investigate the effects of the covalent bonding on the signal generated in such sensors, wafer-wafer bonded silicon-silicon P-N pad diodes have previously been produced.
The behaviour of these test samples is being investigated with transient current technique (TCT) measurements.
In this paper we present an overview of the TCT setup as well as a custom sandwich-type sample holder used for these measurements.
A review of the results presented in a previous paper shows, that the bonded P-N structures show a highly asymmetric depletion behaviour under reverse bias.
IR edge TCT measurements confirm that only the P-side of the samples is being depleted.
}


\begin{document}

\maketitle
\flushbottom

\noindent In this paper we follow-up on our previous investigation on low temperature covalent wafer-wafer bonded silicon sensor diodes as presented in~\cite{wuthrich_depletion_2022}.
The initial results published in~\cite{wuthrich_depletion_2022} highlighted two unexpected findings: (1) a charge collection which increased with bias voltage as expected, up to a maximum at $V_{Bias} \approx 90~\text{V}$, but decreased for $V_{Bias} \geq 90~\text{V}$, which was not understood, and (2) a highly asymmetric depletion pattern where symmetric depletion had been expected.
In this publication we review and clarify both of these findings.
Following a short introduction into the subject (section~\ref{sec:introduction}) and a more detailed description of the TCT setup used (section~\ref{sec:tct_setup}), we show that (1) was caused due to instrumental effects (section~\ref{sec:review_results}).
We correct these effects and show new measurements with the expected charge collection behaviour.
Finally we confirm finding (2) via direct measurements of the depletion profile (whereas our conclusion in~\cite{wuthrich_depletion_2022} was from indirect evidence) and we discuss the implications of this result on the behaviour of the bonding interface (section~\ref{sec:edge_tct_results}).

\section{Introduction}
\label{sec:introduction}

We are investigating the use of low-temperature covalent wafer-wafer bonding in the fabrication of future particle detectors. Wafer-wafer bonding enables the creation of hetero structures by bonding different types of semiconductors (for example GaAs to Si) together. As the bonding is carried out at or near room temperature, it allows to bond fully processed CMOS wafers.
This positions low-temperature wafer-wafer bonding as an alternative to bump-bonding, which is commonly employed in detector fabrication.

X-ray imaging sensors greatly benefit from the use of hetero-structure based detectors, especially in the domain of medical imaging.
In the X-ray energy range employed in medical imaging, the interaction of X-ray photons with matter is dominated by the photo-electric effect and therefore efficient X-ray imaging detectors need to ensure a high interaction probability of photons with matter.
For X-ray energies above 10-20 keV this is commonly achieved by making use of high-Z materials as the interaction part of a detector (such as GaAs, CdZnTe or SiGe).
As it is not economically feasible to implement the highly integrated readout electronics of imaging detectors in these high-Z materials, such sensors are usually paired with a front-end ASIC implemented in a standard CMOS process via bump-bonding~\cite{locker_single_2004} usually carried out on a per chip basis.
It is a complex process with non negligible costs, also due to the limited yield of the process itself.
Further, the additional capacitance and resistance from the connection between sensor and readout ASIC introduces additional noise~\cite{atlas_collaboration_technical_2017}\cite{yarema_advances_2014}.

Low-temperature covalent wafer-wafer bonding is a novel and alternative approach to create hetero-structure particle detectors.
In this approach the covalent bonding process is used to directly fuse the readout ASIC fabricated in a CMOS process with the high-Z sensor layer.
The resulting structure resembles a monolithic detector, but made up of two different materials.
Compared to bump-bonding the low-temperature covalent bonding is carried out on a wafer basis.

The principle of low-temperature covalent wafer-wafer bonding is presented in more detail in~\cite{wuthrich_depletion_2022} and is summarized here.
The bonding process is carried out in ultra-high vacuum where the surfaces of each wafer to be bonded are sputter cleaned via an argon beam or argon plasma.
This sputter cleaning removes surface contaminations as well as the thin native oxide layer and leaves behind activated crystal bonds.
When the two wafer surfaces are brought into contact under moderate pressure at room temperature these activated bonds spontaneously react leading to a high bond strength.
No high temperature annealing is necessary which make this bonding process CMOS compatible.

As discussed in~\cite{wuthrich_depletion_2022} the sputter cleaning affects the crystalline structure at the surface of the two wafers to be bonded and leads to a thin amorphous layer at the bonding interface.
This (non passivated) amorphous layer introduces non-ideal behaviour in the fabricated detector, especially due to the high density of crystal defects.
In order to investigate this non-ideal behaviour we fabricated simple P-N pad diodes by bonding a P-type wafer to an N-type wafer, creating a P-N junction at the bonding interface.
The fabrication process is documented in~\cite{wuthrich_depletion_2022}.
The P and N wafers employed are high-resistivity float-zone wafers as commonly used in the fabrication of particle detectors.
Details about the used wafers can be found in table~\ref{tab:wafer_spec}.

\begin{table}
  \caption{Specifications of the wafers used for this production run. From~\cite{wuthrich_depletion_2022}.}
  \label{tab:wafer_spec}

  \vspace{0.3cm}

  \resizebox{\textwidth}{!}{
    \begin{tabular}{l|rrrrr}
      \textbf{Type} & \textbf{Orientation} & \textbf{Size} & \textbf{Thickness} & \textbf{Resistivity} & \textbf{Doping} \\ \hline
      P (Boron) & <100> & $100 ~ \text{mm}$ & $500 \pm 7 ~ \mu\text{m}$ & $> 10 ~ \text{k}\Omega\text{cm}$ & $< 2\cdot10^{12} ~ \text{cm}^{-3}$ \\
      N (Phosphorous)& <100> & $100 ~ \text{mm}$ & $490 \pm 8 ~ \mu\text{m}$ & $7 - 10 ~ \text{k}\Omega\text{cm}$ & $(4 - 6) \cdot10^{11} ~ \text{cm}^{-3}$ \\
    \end{tabular}
  }
\end{table}

\begin{figure}
  \centering
  \includegraphics[width=0.7\textwidth]{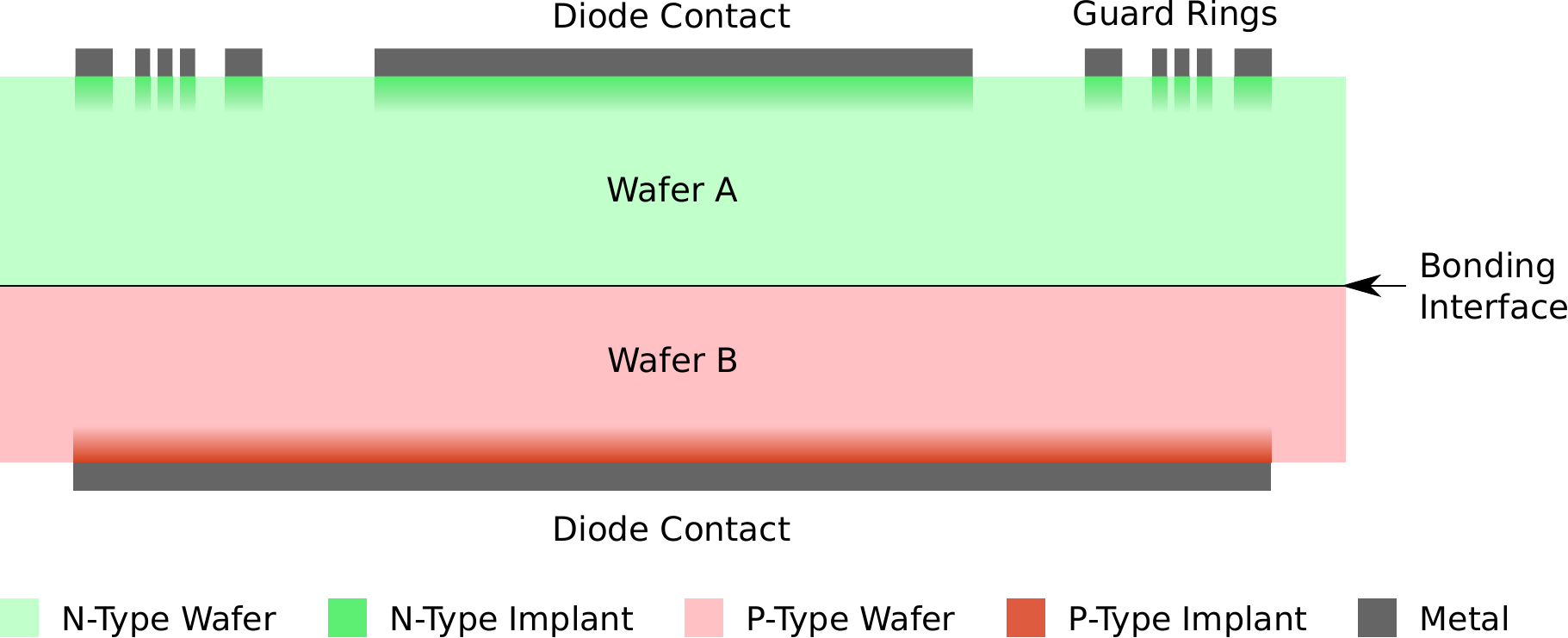}

  \caption{Schematic view of the cross-section of the fabricated test structures. A high resistivity N-type silicon wafer is bonded to a high resistivity P-type wafer. Each wafer has high-dose implants at the surface in order to ensure good ohmic conduction between the silicon substrate and the metal contacts. Image not to scale. From~\cite{wuthrich_depletion_2022}.}
  \label{fig:diode_crosssection}
\end{figure}

The investigations presented in this paper are based on measurements of pad diodes with a size of 5.6x5.6 mm.
The diodes have a $1~\text{mm}$ diameter opening in the front and back-side metallizations, allowing for TCT measurements via the diode contacts.
The samples employed for the measurements presented in this paper are the same as in~\cite{wuthrich_depletion_2022}.
Figure~\ref{fig:diode_crosssection} shows the schematic cross-section of the bonded diode sample.

\section{Transient Current Technique Setup}
\label{sec:tct_setup}

The transient current technique (TCT) measurements were carried out on a setup based on the \textit{Large Scanning TCT} setup provided by Particulars~d.o.o.~\cite{particulars_doo_particulars_2021}.
The main parts of the setup provided by Particulars are the diode laser sources and drivers, the optical system, the electrical front-end (high-gain amplifier, bias-tee and high voltage supply filter) and mechanical system (\textit{x-y}~sample translation stage and \textit{z}~focus translation stage).
For different types of measurements we employ a red laser ($\lambda_{Red} = 660~\text{nm}$) and an infrared (IR) laser ($\lambda_{IR} = 1064~\text{nm}$).
In addition, the following main components are used:
\begin{itemize}
  \item \textbf{HV supply:} Keithley 2410 SMU ($\pm 1000~\text{V}$)~\cite{tektronix_models_2022}
  \item \textbf{Waveform acquisition:} Lecroy Waverunner 8104 Oscilloscope (1GHz -- 20Gsps) \cite{teledyne_lecroy_waverunner_2022}
  \item \textbf{Control and data acquisition (DAQ):} Linux based computer running Debian \cite{software_in_the_public_interest_inc_debian_2022}
  \item \textbf{Stage temperature control:} Lauda LOOP L 100 \cite{lauda_dr_r_wobser_gmbh__co_kg_circulation_2022} 
  \item \textbf{Amplifier supply:} Aim TTi PLH250-P \cite{aim_and_thurlby_thandar_instruments_plh_2022} 
\end{itemize}
All devices are connected to the control computer via USB or Ethernet and the entire setup can be operated remotely.
The only manual operations necessary are switching from the red to the IR laser (and vice-versa) as well as swapping the devices under test (DUT).
The control and data acquisition is executed via a custom Python-based framework.
All devices are automatically configured via the control framework, which ensures consistent repeatability of measurements across multiple samples.
The sample temperature regulation is achieved via thermoelectric elements mounted to the TCT sample stage.
The active proportional–integral–derivative (PID) controller uses the temperature measured directly on the sample holder (see section~\ref{sec:sandwich_type_holder}) as a control input.
The Lauda L 100 circulating thermostat extracts the excess thermal energy from the thermoelectric elements.

\subsection{Sandwich-Type Sample Holder}
\label{sec:sandwich_type_holder}

\begin{figure}
  \centering
  \begin{subfigure}[c]{0.45\textwidth}
    \includegraphics[width=\textwidth]{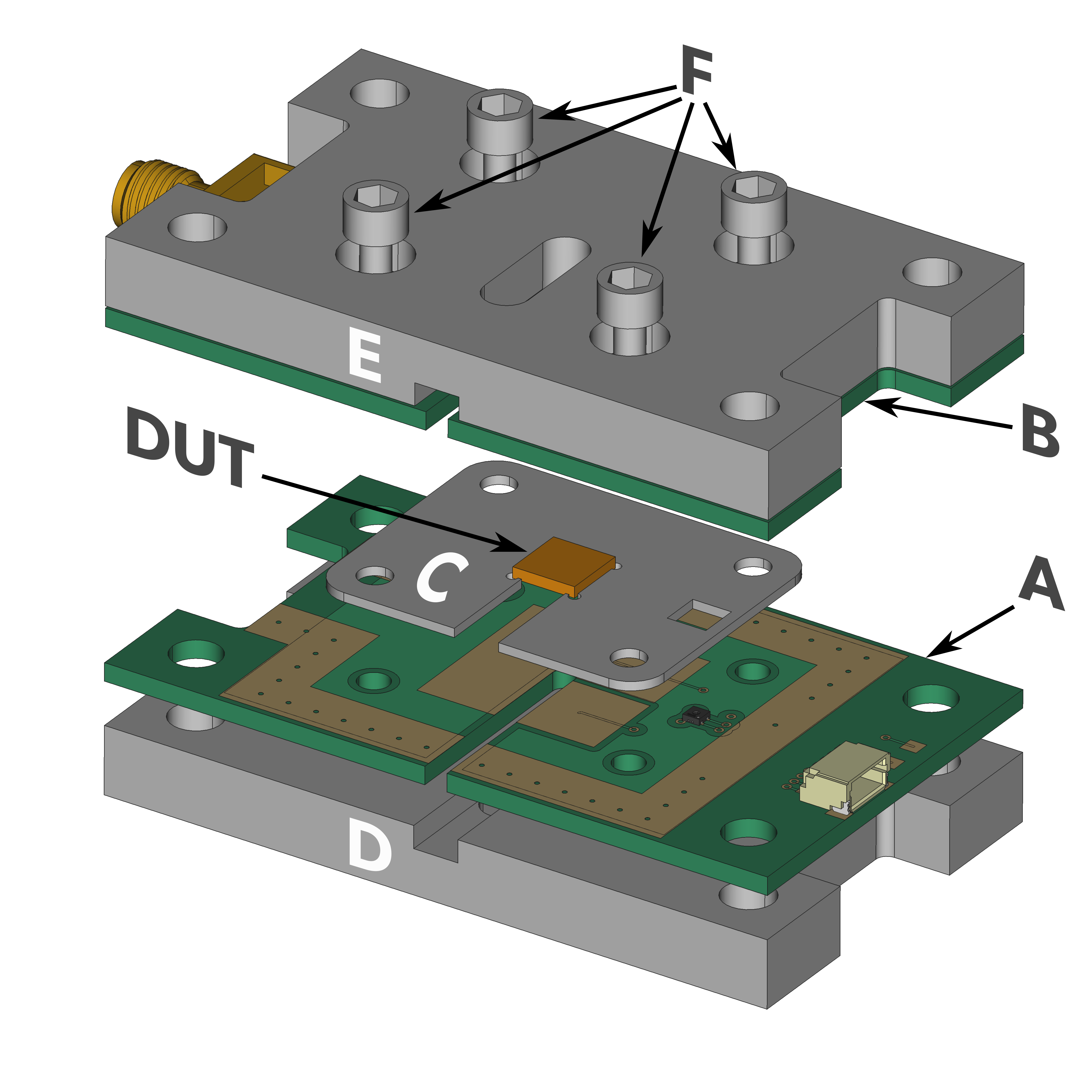}
  \end{subfigure}%
  \hfill
  \begin{subfigure}[c]{0.5\textwidth}
    \includegraphics[width=\textwidth]{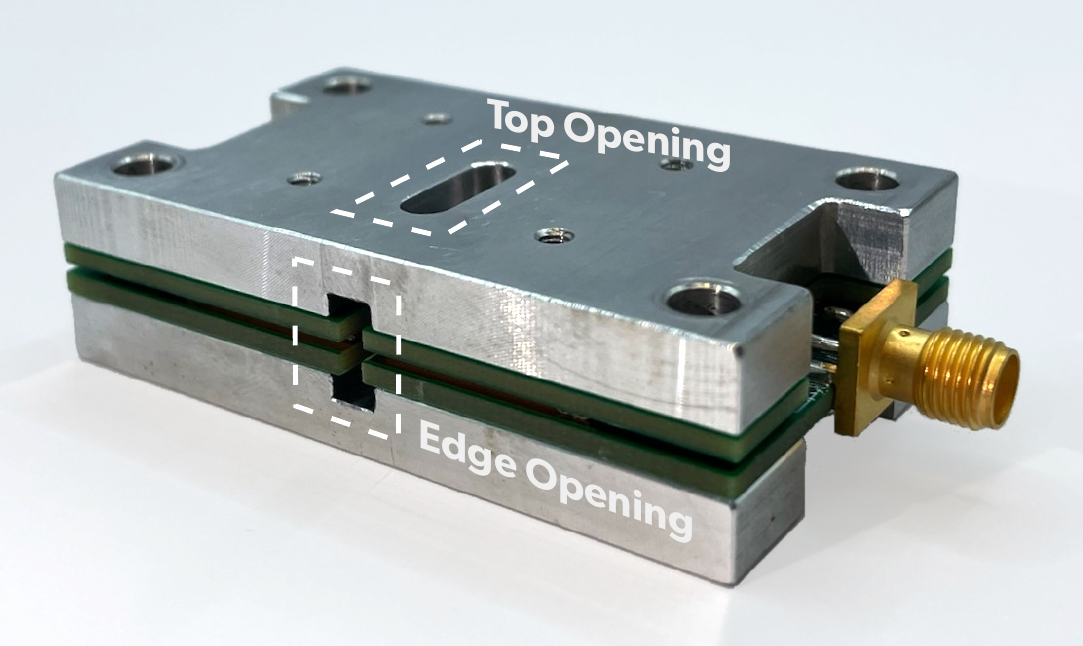}
  \end{subfigure}
  \caption{Left: Exploded view of the sandwich-type sample holder, showing the diode sample~(DUT), PCBs~(A and~B), the 3D-printed spacer piece~(C), metal support pieces (D and~E) and tensioning screws~(F)~--~Right:~Image of the assembled sample holder, indicating the openings used for top and edge TCT.}
  \label{fig:holder_explosion}
\end{figure}

Sample holders used for TCT measurements usually require the DUT to be wire bonded to the holder.
In contrast, we built a sandwich-type holder which clamps the DUT in between two custom PCBs, without the need for wire bonding.
The contact to the DUT is made by bringing into contact the metallization on the surfaces of the DUT with the PCB top metal layer.
Figure~\ref{fig:holder_explosion}~(left) shows an exploded view of the sample holder assembly.
Figure~\ref{fig:holder_explosion}~(right) shows a picture of the assembled sample holder.

The diode sample (DUT) is laterally held in place by a 3D printed spacer piece marked~C in figure~\ref{fig:holder_explosion}~(left).
The spacer piece is slightly thinner than the DUT in order to ensure that the contact pressure is fully applied to the DUT.
The two PCBs~(marked A and~B) enclose the DUT from both sides.
Each PCB is supported by a metal support piece on the back side~(marked D and~E).
The screws marked F are used to tighten the sandwich to ensure good electrical contact between the DUT and the PCBs~\cite{timsit_electrical_1998}.
It is important that the support pieces are made of a stiff material (i.e. aluminium) which does not allow for any flex, as any bending of the PCBs leads to a bad contact.

\begin{figure}
  \centering
  \includegraphics[width=\textwidth]{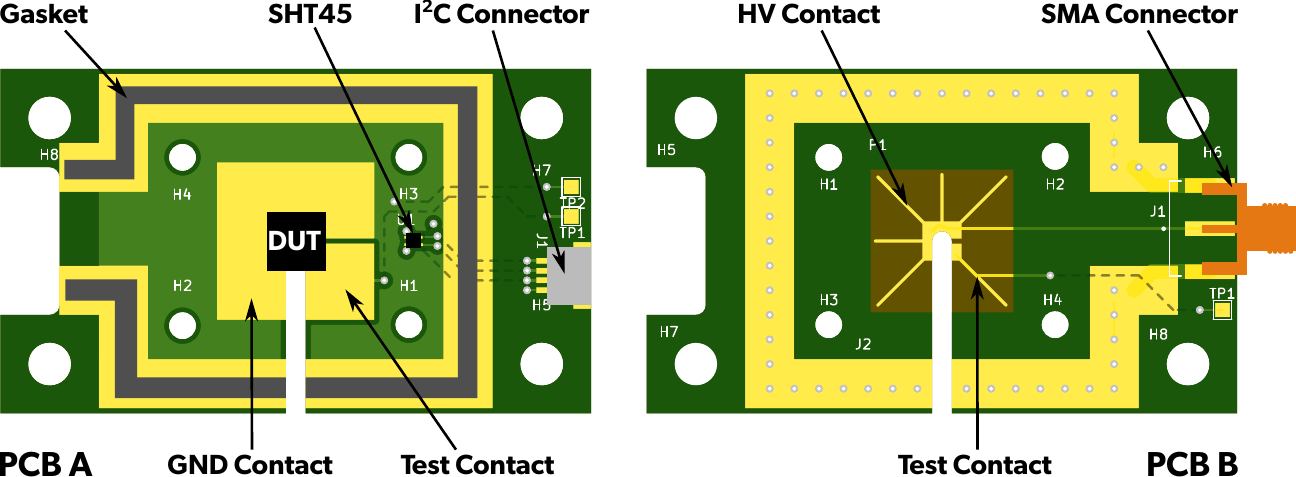}

  \caption{Annotated top layer view of the two PCBs used in the sandwich-type sample holder. PCB~A includes temperature and humitdy readout via a SHT45 sensor from Sensirion. PCB~B shows the specially designed low capacitance contact and a matched $50~\Omega$ transmission line up to the SMA connector.}
  \label{fig:pcb_layout}
\end{figure}

The PCBs have a special metal patterning which allows to measure the contact resistance between PCB and DUT using a common hand-held multimeter.
Figure~\ref{fig:pcb_layout} shows the layout of the two PCBs.
The metal region which is in contact with the DUT is split in two isolated parts which are individually connected to test points on the PCB.
In figure~\ref{fig:pcb_layout} these two parts are marked \textit{GND Contact} and \textit{Test Contact} for PCB~A, and \textit{HV Contact} and \textit{Test Contact} for PCB~B.
After installation of a sample diode, these isolated parts are shorted together by the surface metallization of the DUT.
By measuring the resistance between the test points, one can probe the contact resistance between DUT and PCB and thus verify that a good contact has been established.
The PCBs were fabricated with an electroless nickel immersion gold (ENIG) surface plating.
The thin gold surface coating does not only resist oxidation, but also provides improved surface planarity, especially compared to a HASL type surface finish~\cite{lentz_how_2018}.

PCB~A in figure~\ref{fig:pcb_layout} creates the ground (GND) connection with the P-side of the diode samples.
A Sensirion SHT45 sensor~\cite{sensirion_ag_sht45_2022} read out via a dedicated I2C connection is mounted on the PCB.
This enables temperature and humidity measurements in close proximity to the DUT.
The GND potential is supplied from the PCB~B to PCB~A via compressible and conductive gaskets\footnote{For example as produced by Würth Elektronik~\cite{wurth_elektronik_eisos_gmbh__co_kg_we-lt_2020}.}.
These gaskets are mounted on the area marked accordingly on PCB~A.
Once the sandwich is assembled the gaskets form a contact with the corresponding area on PCB~B.
Together with dedicated GND planes on each PCB, the gaskets form a Faraday cage around the DUT.
This greatly reduces the noise pickup and enables better data acquisition within the Particulars TCT setup, which by itself does not provide sufficient shielding.

The PCB~B in figure~\ref{fig:pcb_layout} creates the high-voltage (HV) connection with the N-side of the DUT.
HV and GND are supplied to the PCB~B via the edge mount SMA connector.
The signal is also collected via the N-side / HV contact.
One challenge of such a sandwich based sample holder is the reduction of the parasitic capacitance introduced in parallel with the DUT.
The various metal layers of the two PCBs act as parallel plate capacitors.
In order to reduce these added parasitic capacitances, one can either reduce the surface area or move the surfaces farther apart\footnote{According to the capacitance calculated via the infinite plane approximation $C = \epsilon \frac{A}{d}$, with $A$ the plate area, $d$ the plate separation and $\epsilon$ the permittivity.}.
Accordingly, the GND planes which form the Faraday cage have been moved to the outermost layers of the 4-layer PCB stacks.
To reduce the parasitic capacitance of the metal layers in contact with the DUT the area of the metal contact on PCB~B was minimized, taking the form of the star-like structure in the centre of the PCB.
The parasitic capacitance of a fully assembled sample holder is $C_{para} \approx 8.2~\text{pF}$ as measured with a Keysight~U1733C LCR-meter at 100kHz~\cite{keysight_technologies_u1730c_2018}.
This value neither includes the capacitance of the DUT, nor the capacitance of the electronics front-end.
The HV contact traces are designed to have a $50~\Omega$ impedance matching when the PCBs are fabricated using the \textit{JLC7628} 4-layer impedance control stackup from JLCPCB~\cite{jlcpcbcom_controlled_2022}.

As marked in figure~\ref{fig:holder_explosion}~(right), there are openings in the metal support structures as well as the PCBs which allow for laser illumination via the N and P-side of the DUT, as well as via the edge of the DUT.
This enables the sample holder to be used for both top/bottom and edge TCT measurements without remounting the DUT.
Overall we identify the following advantages of this type of sandwich-type holder:
\begin{itemize}
  \item No specialized tools (e.g. wire boding) are necessary for mounting a sample.
  \item Mounting a new sample is very quick and the sample holder can be reused without any issues.
  \item The sample holder provides immediate shielding to noise pickup.
\end{itemize}
Compared to traditional sample holder using wire bonding there are certain disadvantages:
\begin{itemize}
  \item The sandwich structure leads to higher parasitic capacitances.
  \item It can only be used for pad diodes with reasonably large metal contacts.
  \item It can not be used with samples which are sensitive to pressure.
  \item There is no trivial way of separately connecting guard rings.
\end{itemize}

\section{Review of the results in~\cite{wuthrich_depletion_2022}}
\label{sec:review_results}

In a previous publication~\cite{wuthrich_depletion_2022} we reported on the initial TCT results obtained with bonded diode samples.
We investigated the depletion behaviour of the samples via top TCT measurements from both the N- and P-side, using red and IR lasers.
Based on red laser measurements, we could show that the P-side of our structures can be fully depleted at a bias voltage $V_{Red} \approx 82~\text{V}$.
On the other hand, no signal could be observed from red laser illumination on the N-side at $V_{Bias}~\text{up to}~650~\text{V}$.
As the N and P-side have a similar doping concentration, one would naively expect a symmetric depletion behaviour of the two sides.
The fact that no signal was observed from illuminating the N-side with the red laser hinted towards an asymmetric depletion behaviour.
Using IR laser illumination we observed a maximum charge collected at a bias voltage $V_{IR} \approx 90~\text{V}$ very close to $V_{Red}$.
But it was also observed, that the integrated signal collected under IR laser illumination decreased for $V_{Bias} > V_{IR}$.
This can not be explained by any effect usually occurring in reverse biased sensors.
Increasing $V_{Bias}$ either leads to an increase in collected signal, due to additional charges being deposited in the depletion region, or leads to a constant collected signal once the maximum extent of the depletion zone has been reached.

\begin{figure}
  \centering
  \includegraphics[width=0.75\textwidth]{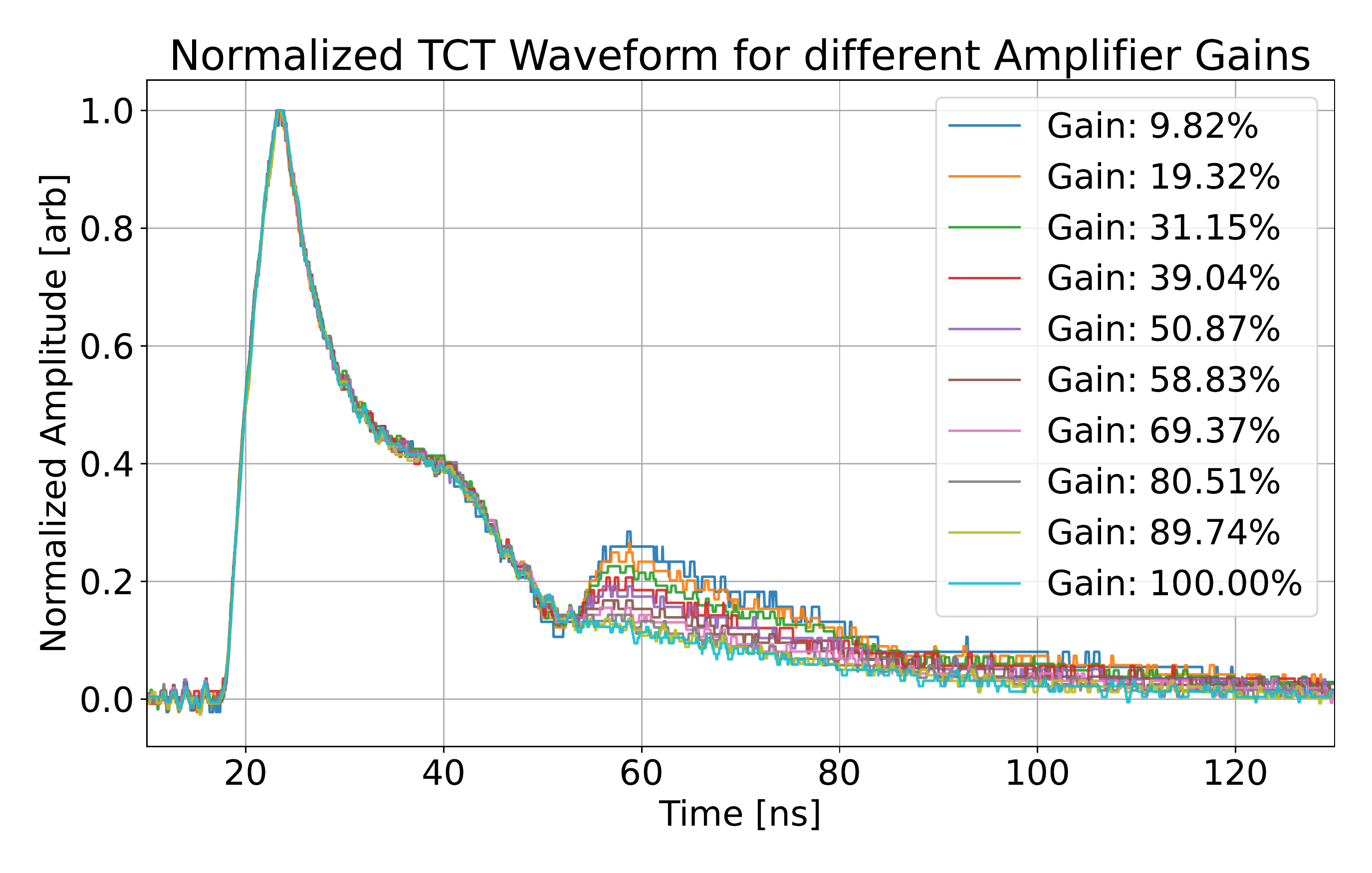}

  \caption{Normalized recorded TCT pulses in function of the relative amplifier gain setting. Gain values below 100\% lead to the appearance of a second bump in the TCT curves causing a non-linear distortion of the collected signal. All recorded pulses shown in this figure have an absolute amplitude of less than $200~\text{mV}$. The initial peak at $20~\text{ns}$ corresponds to electron movement and the plateau until $40~\text{ns}$ to hole movement.}
  \label{fig:pulse_distortion}
\end{figure}

Investigations showed, that this unexplained observation is caused by non-linearities in the electronics front-end of the TCT setup.
For the measurements presented in~\cite{wuthrich_depletion_2022} the Particulars AM-02A amplifier was operated with a supply voltage of ca. $7~\text{V}$ which, according to its specifications, results in ca. 50\% of the nominal gain~\cite{particulars_doo_particulars_2017}.
Our investigation showed, that operating our amplifier at less than 100\% nominal gain leads to a non-linear secondary bump in the TCT time domain signal.
Figure~\ref{fig:pulse_distortion} shows amplitude normalized TCT pulses recorded with different amplifier gain values.
It is clearly visible that the secondary bump (at ca. 60ns) is well correlated with the amplifier gain.
The origin of the secondary bump is not known, but operating the amplifier at 100\% nominal gain leads to a clean signal.

\begin{figure}
  \centering
  \includegraphics[width=\textwidth]{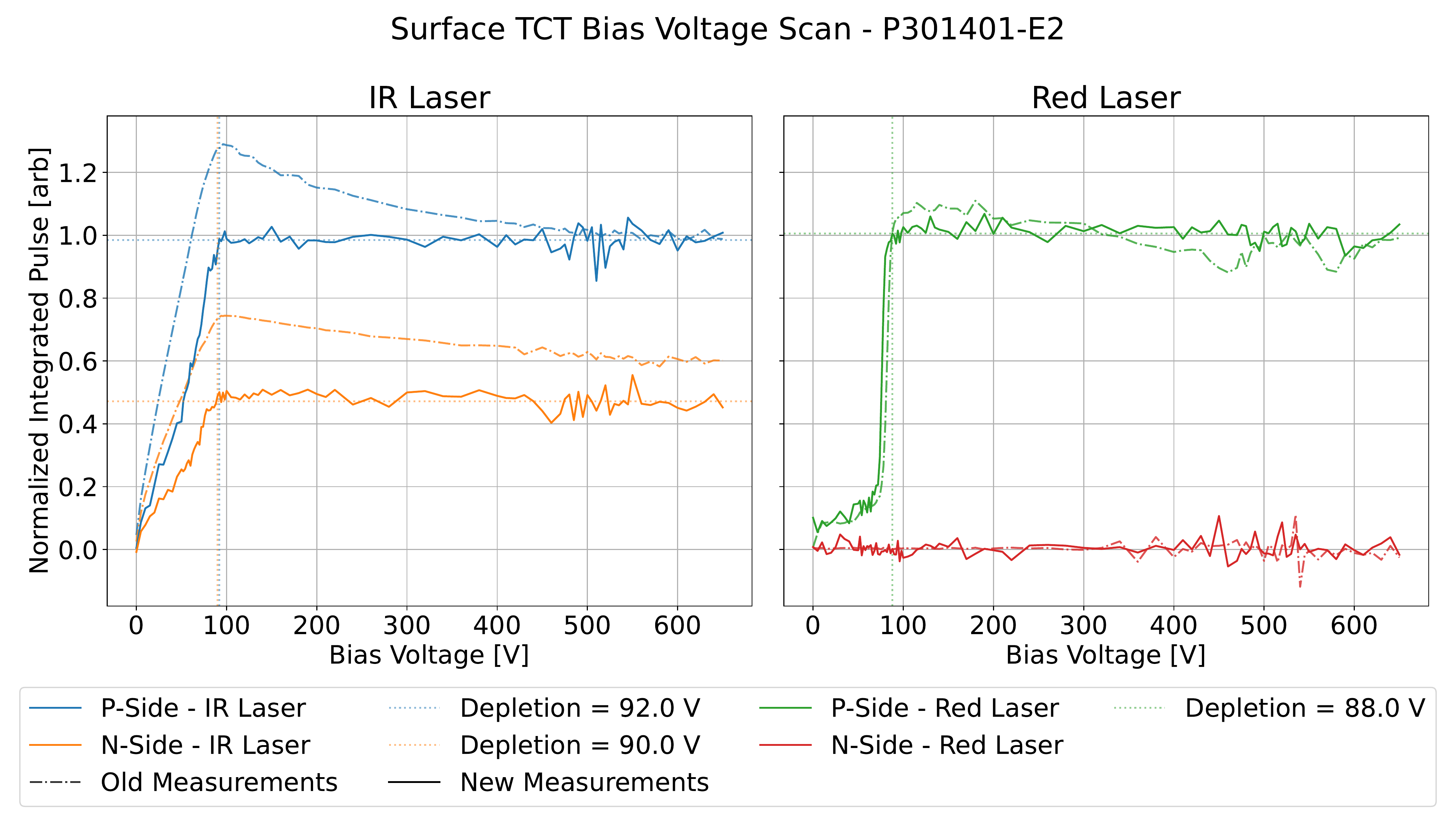}

  \caption{Comparison of the integrated signal curves before and after changing the amplifier gain. The old measurements were done with a relative amplifier gain less than 100\% which led to non-linear effects in the time domain TCT curves.}
  \label{fig:reviewed_pulse_integral}
\end{figure}
The investigation carried out in~\cite{wuthrich_depletion_2022} was repeated with the improved amplifier gain settings.
Figure~\ref{fig:reviewed_pulse_integral} shows the resulting integrated pulse curves, as well as the previously measured curves as a reference.
In the new measurements the decrease in collected charge from IR TCT at high voltages has disappeared.
We therefore attribute this previous observation to the non-linear behaviour of the amplifier operated below 100\% nominal gain.
Accordingly, any future measurements were carried out with a gain setting of 100\%.

In light of these changes we review the argumentation presented in section~3.1 of~\cite{wuthrich_depletion_2022}.
We again define $V_{Red} \approx 88~\text{V}$ to be the bias voltage at which the signal observed under red laser illumination of the P-side is saturating.
This value indicates the bias voltage necessary for fully\footnote{Due to the presence of the highly doped ($\approx 5 - 10 \cdot 10^{19}~\text{cm}^{-3}$) contact region, there is always a thin undepleted layer of ca. $1~\mu\text{m}$ at the P-side surface.} depleting the P-side of our structure.
Similarly we define $V_{IR,P} \approx 92~\text{V}$ and $V_{IR,N} \approx 90~\text{V}$ to be the bias voltages at which the signal observed under P-side, resp. N-side IR laser illumination saturates.
Under consideration of the uncertainty introduced to these values due to the signal noise, we have to assume that one single maximum depletion voltage exists.
Accordingly we can conclude that the depletion of our structure stops, as soon as the P-side is fully depleted.
The bulk of the N-side is homogeneous and therefore does not have any structure which would impose a limit to the depletion on this side of the structure.
We therefore arrive to the hypothesis, that only the P-side is depleting under reverse bias, and that no depletion of the N-side bulk occurs.
Under this assumption the electrons necessary for depleting the P-side must be provided by the non-ideal bonding interface.
Indeed, the results of the edge TCT measurements presented in the next section will confirm this interpretation.

\section{IR Edge TCT Measurements}
\label{sec:edge_tct_results}

In order to better understand the depletion behaviour of our samples, edge TCT measurements were carried out for a total of six samples.
The six samples were taken from three different bonded wafer pairs.
All have the same P-N bonded structure, but exhibit variations in the P and N wafer bulk doping concentrations.
The samples are summarized in table~\ref{tab:edge_samples}.
\begin{table}
  \caption{List of diode samples used for the edge TCT measurements.}
  \label{tab:edge_samples}

  \centering
  \vspace{0.3cm}
  \begin{tabular}{lll|rr}
    \textbf{Bond-ID} & \textbf{P-Wafer} & \textbf{N-Wafer} & \textbf{Sample-IDs} & \textbf{Sample-Size} \\ \hline
    P301401 & 301 & 401 & \textit{P301401-E2} and \textit{P301401-E12} & $5.6 \times 5.6 ~\text{mm}^2$ \\
    P302402 & 302 & 402 & \textit{P302402-E2} and \textit{P302402-E12} & $5.6 \times 5.6 ~\text{mm}^2$ \\
    P303403 & 303 & 403 & \textit{P303403-E2} and \textit{P303403-E12} & $5.6 \times 5.6 ~\text{mm}^2$ \\
  \end{tabular}
\end{table}

\begin{figure}
  \centering
  \includegraphics[height=0.35\textheight]{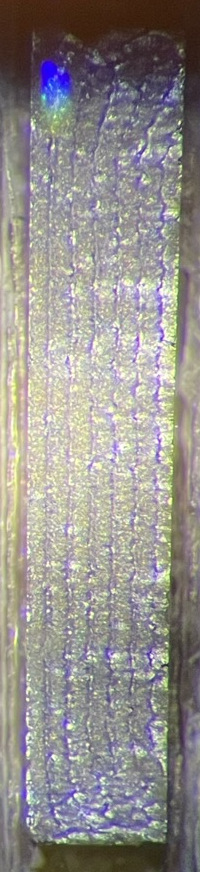}
  \includegraphics[height=0.35\textheight]{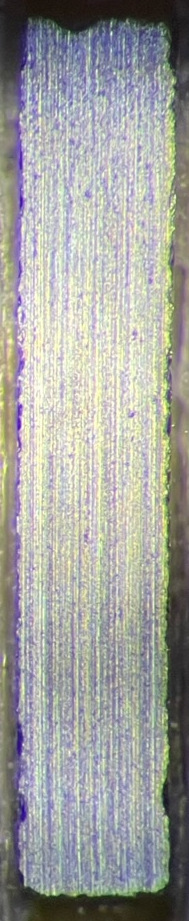}
  \includegraphics[height=0.35\textheight]{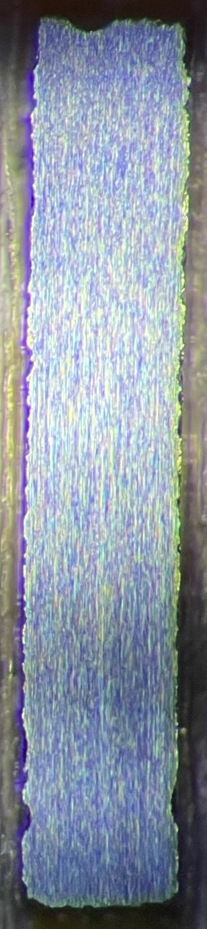}
  \includegraphics[height=0.35\textheight]{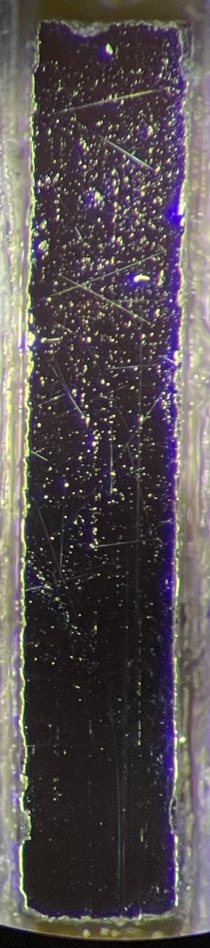}
  \includegraphics[height=0.35\textheight]{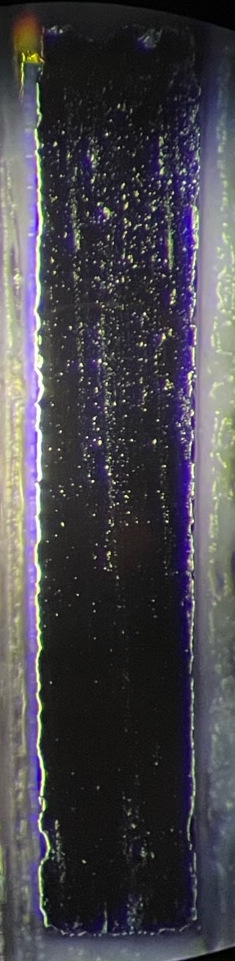}

  \caption{Microscopic view of a sample edge in various stages of polishing. From left to right: unpolished edge after laser dicing; after polishing with $30~\mu\text{m}$ grain lapping paper; after polishing with $3~\mu\text{m}$ grain lapping paper; after polishing with $0.02~\mu\text{m}$ grain lapping paper; after polish with diamond paste using a cotton swab. The samples were illuminated with a ring light. Thus the more polished the surface is, the less light is diffusively reflected into the objective of the microscope and the darker the edge appears in the image.}
  \label{fig:edge_polish}
\end{figure}

Laser dicing was used to cut the bonded wafers into individual pad diodes.
The diced samples have rough edges with a periodic pattern which originates from the multiple passes necessary for the laser to fully cut through the wafer.
This edge surface pattern leads to variations in absorption of laser light, as well as reflection and scattering of light in multiple directions.
It is therefore primordial to polish the edge surface before any edge TCT measurement.
Our samples were polished in a multistep process, first with lapping paper with grain sizes of $30~\mu\text{m}$, $3~\mu\text{m}$ and $0.02~\mu\text{m}$.
The fine polish was done using a diamond polishing paste with a grain size of $0.1~\mu\text{m}$ applied via cotton swabs~\cite{feindt_edge-tct_2017}.
Figure~\ref{fig:edge_polish} shows an example edge after the various polishing steps.

\begin{figure}
  \centering
  \includegraphics[width=\textwidth]{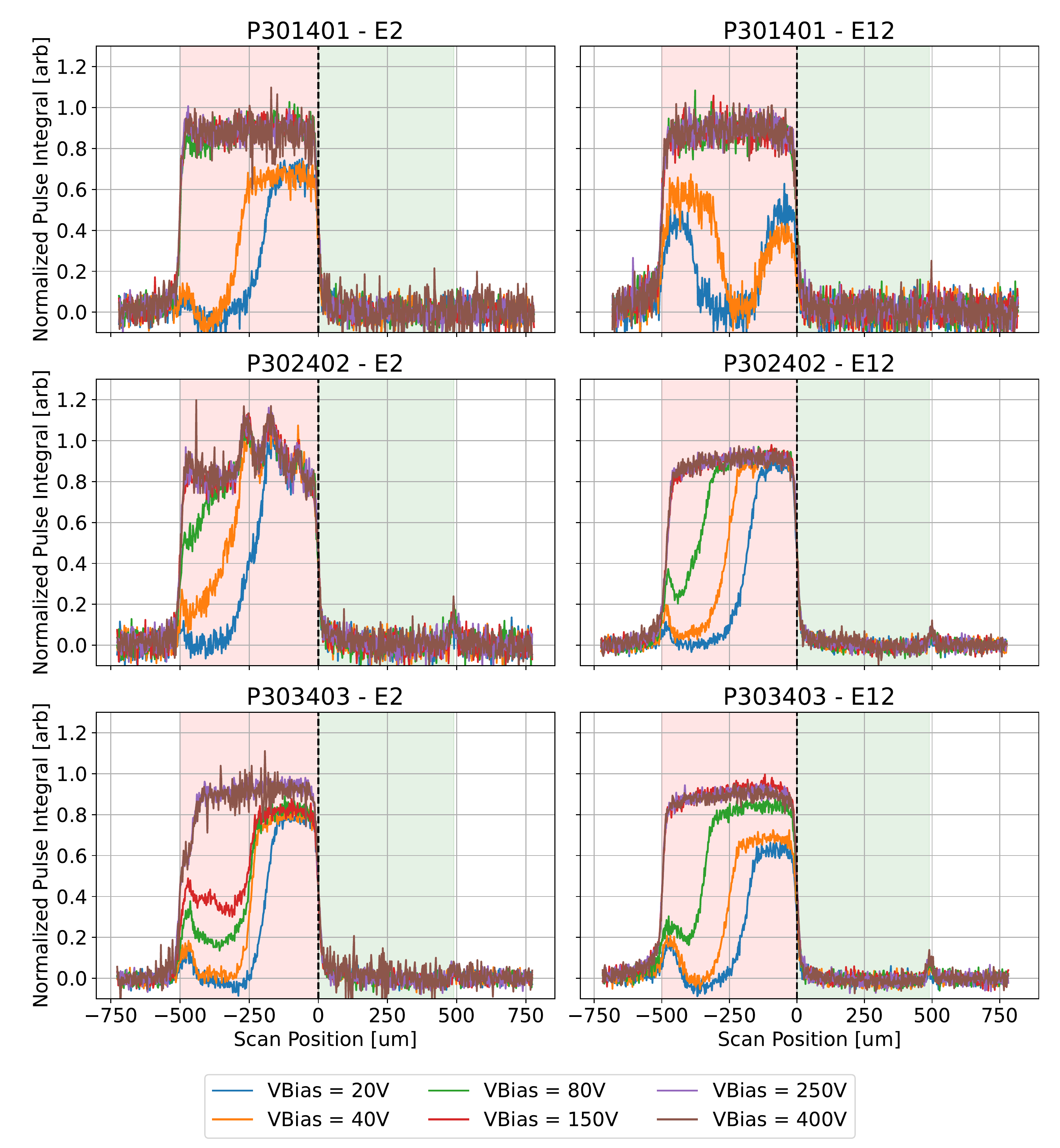}

  \caption{Bias dependent signal of the measured diode samples as a function of the position along the diode cross-section. The bonding interface is located at $x_{B} \approx 0~\mu\text{m}$. The curves are multiplied with a normalization factor fixed for each sample. The P-side is shown with a red and the N-side with a green background.}
  \label{fig:bias_scan_plots}
\end{figure}

\begin{figure}
  \centering
  \includegraphics[width=\textwidth]{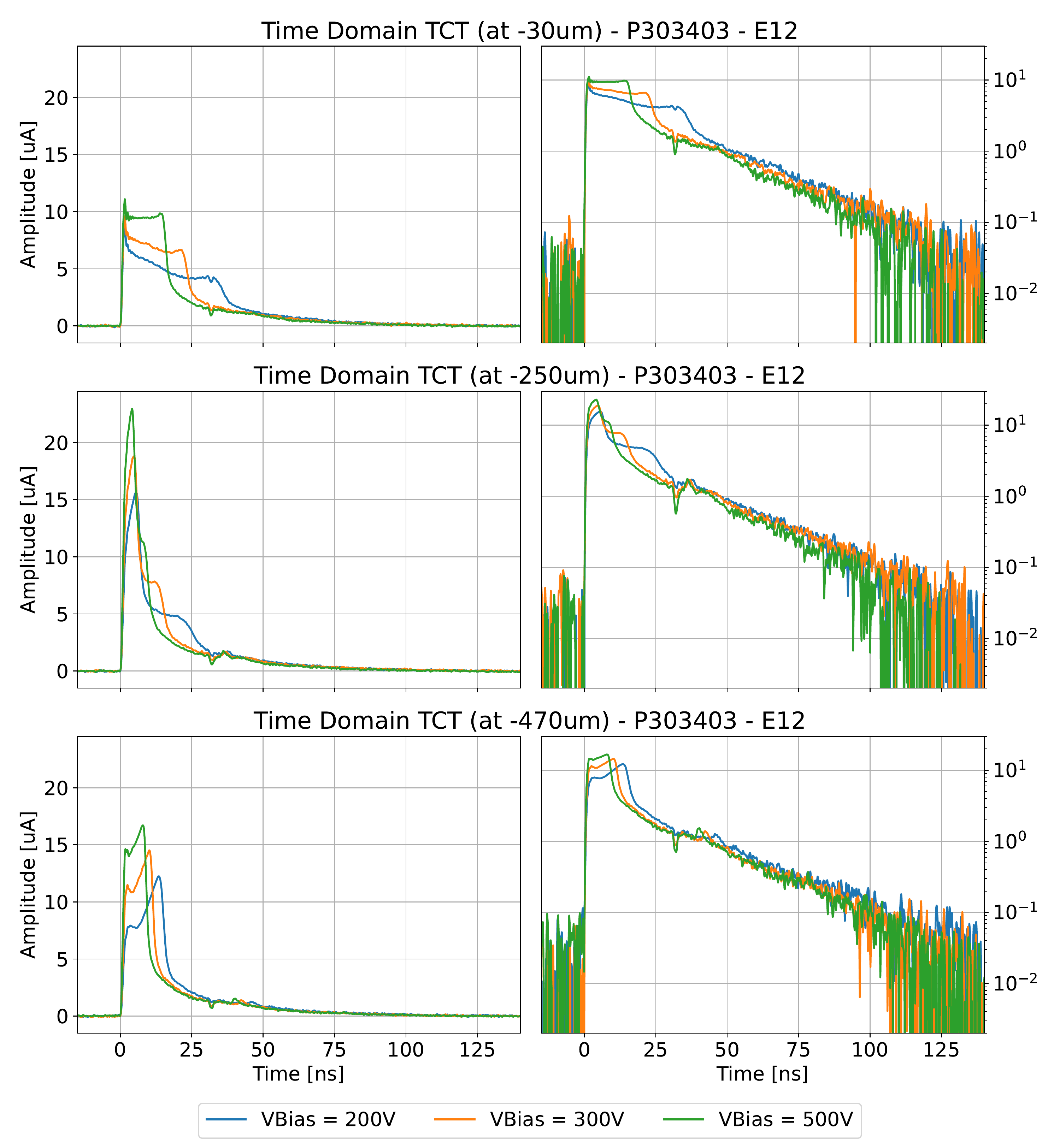}

  \caption{Time domain TCT curves recorded at different bias voltages at $-30~\mu\text{m}$, $-250~\mu\text{m}$ and $-470~\mu\text{m}$ from the bonding interface (left: linear scale -- right: logarithmic \textit{y}-scale). The curves at $-30~\mu\text{m}$ mainly show hole drift and the curves at $-470~\mu\text{m}$ mainly electron drift. The artefact at ca. $30~\text{ns}$ is due to signal reflections at the front-end electronics.}
  \label{fig:timedomain_curves}
\end{figure}

The measurements were performed using the IR laser ($\lambda = 1064~\text{nm}$).
The focus of the laser was evaluated with a standard knife edge scan for each sample individually~\cite{suzaki_measurement_1975}.
The focused laser beam has a Gaussian beam profile with a measured standard deviation of $\sigma \approx 10~\mu\text{m}$.
Standard TCT scans were executed by moving the laser across the polished edge (from bottom to the top in figure~\ref{fig:diode_crosssection}); we denote this a scan along X.
Scans were carried out across the entire diode structure (P- and N-side) with a step size of $\Delta{}x = 2~\mu\text{m}$ along X.
In all plots to be presented, the P-side surface is at~$x_{P} \approx -500~\mu\text{m}$, the bonding interface is at~$x_{B} \approx 0~\mu\text{m}$ and the N-side surface at~$x_{N} \approx 490~\mu\text{m}$.\footnote{The P-type (N-type) wafers have a nominal thickness of $W_{P} = 500~\mu\text{m}$ ($W_{N} = 490~\mu\text{m}$) after CMP polishing (last processing step before bonding). The variation in thickness of the wafers after polishing is ca. $\pm 5~\mu\text{m}$.}
In the following plots the P-side is presented with a light red background and the N-side with a light green background.
The interface is indicated with a black dotted line.
X-scans were carried out at different reverse bias voltages in the range $|V_{Bias}| \in [2, 600]~\text{V}$.
The oscilloscope records the TCT curves as a voltage signal $U_{V_{Bias}}(x, t)$.
Given the input resistance $R_{In} = 50~\Omega$ of the frontend electronics as well as the gain $A = 53~\text{dB}$ of the amplifier we can calculate the collected signal charge by integrating over the recorded TCT curves as $Q_{V_{Bias}}(x) = \frac{1}{R \cdot A} \int U_{V_{Bias}}(x, t) dt$\footnotemark.
\footnotetext{For our analysis we integrate the pulses up to $450~\text{ns}$, due to the presence of a long exponential tail (see figure~\ref{fig:timedomain_curves}).}
The results of these bias dependent scans is plotted in figure~\ref{fig:bias_scan_plots}.
The curves presented in this figure are multiplied with a normalization factor, fixed for each sample individually.
This was done in order to allow easier comparison between the samples as the laser intensity (and therefore the measured integrated charge) fluctuated in between different samples.
From these curves it is immediately visible, that no signal is collected from the N-side of the samples.
This implies a zero electric field inside the N-side and accordingly no depletion region extending into this part of the samples.
Consequently we confirm the hypothesis stipulated in section~\ref{sec:review_results}, that only the P-side of our samples is being depleted.

Based on these observations the expected standard behaviour is that the interface acts similar to a highly doped N++ region.
The depletion of the samples starts from the bonding interface and only grows into the P-sides of the samples.
Looking at the plots in figure~\ref{fig:bias_scan_plots} we observe that two of the measured samples show a different depletion behaviour.
The sample \textit{P301401-E12} mainly shows a depletion region which grows from the P-side towards the bonding interface.
The sample \textit{P303403-E2} also shows a partial depletion which starts from the P-side surface.
We attribute this behaviour to accidental / parasitic doping during the processing of the wafers.
Most of the processing was done in the BRNC cleanroom of ETHZ / IBM, which, as a user cleanroom, can not guarantee the same level of cleanliness as professional cleanrooms.
The resulting depletion behaviour of these two samples strongly resembles the behaviour of irradiated sensor samples which exhibit a double peak electric field due to doping inversion~\cite{eremin_origin_2002}.
\begin{figure}
  \centering
  \includegraphics[width=\textwidth]{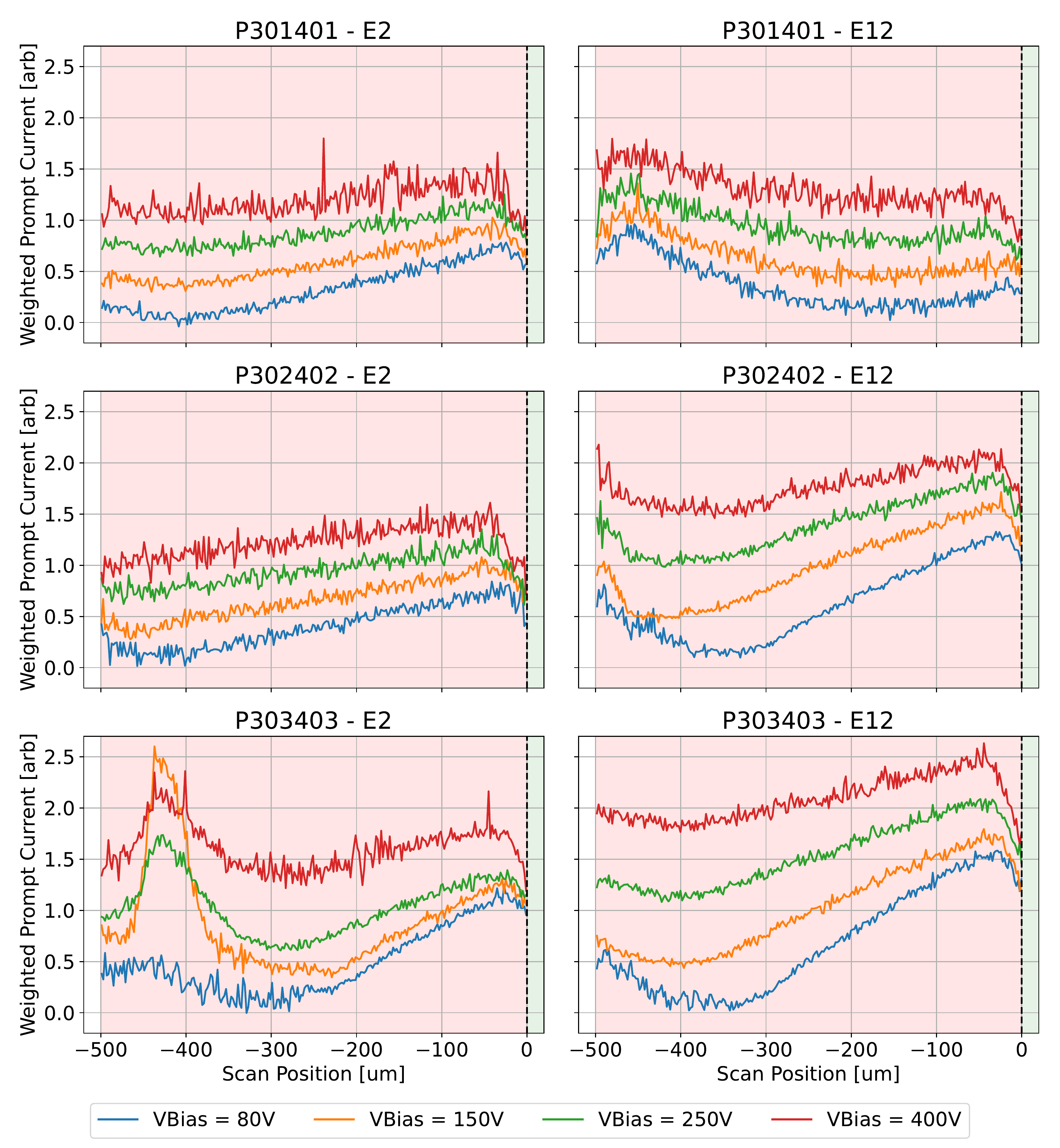}

  \caption{Weighted prompt current curves for all measured samples at various bias voltages. The prompt current curves were calculated according to~\cite{pape_techniques_2022}\cite{kramberger_investigation_2010}. The curves are only calculated within the P-side of the samples.}
  \label{fig:weighted_prompt_plots}
\end{figure}
The same conclusion can be reached when comparing the weighted prompt current curves for all samples, as shown in figure~\ref{fig:weighted_prompt_plots}.
The weighted prompt current curves are proportional to the drift velocity profiles within the diode samples~\cite{kramberger_investigation_2010}.
Given that we expect the depletion region in our samples to start at the bonding interface and growing into the P-side, we also expect the charge drift velocity to gradually increase towards the bonding interface.
The abnormal behaviour of~\textit{P301401-E12} can also be seen in figure~\ref{fig:weighted_prompt_plots}, as the drift velocity profile is inverted from the baseline expectation.
It has a maximum towards the P-side surface and then decreases towards the bonding interface.
Similarly, the unexpected depletion behaviour of~\textit{P303403-E2} can be observed in the weighted prompt current curves, having a strong peak at $x \approx -400~\mu\text{m}$.
We therefore exclude both of these samples from further analysis.

From the curves in figure~\ref{fig:bias_scan_plots} it also appears that~\textit{P302402-E2} shows irregular behaviour, as it exhibits local maxima and minima in collected signal instead of a flat signal response curve.
This behaviour is not due to any irregular intrinsic properties of the sample, but rather due to non-optimal polishing of the sample edge.
The maxima and minima occur because more or less IR light is penetrating into the sample due to the local roughness of the sample edge, leading to a higher (or lower) number of created e$^{-}$/h~pairs.
This is confirmed by the weighted prompt current curves of~\textit{P302402-E2} in figure~\ref{fig:weighted_prompt_plots}, which do not exhibit these local minima and maxima.
As part of the weighted prompt current analysis, the prompt current is divided by the total integrated charge, thus cancelling out any effects which lead to a higher (or lower) number of created e$^{-}$/h~pairs.
From the resulting weighted prompt current curves we can see that sample~\textit{P302402-E2} also shows a standard behaviour.

\begin{figure}
  \centering
  \includegraphics[width=0.75\textwidth]{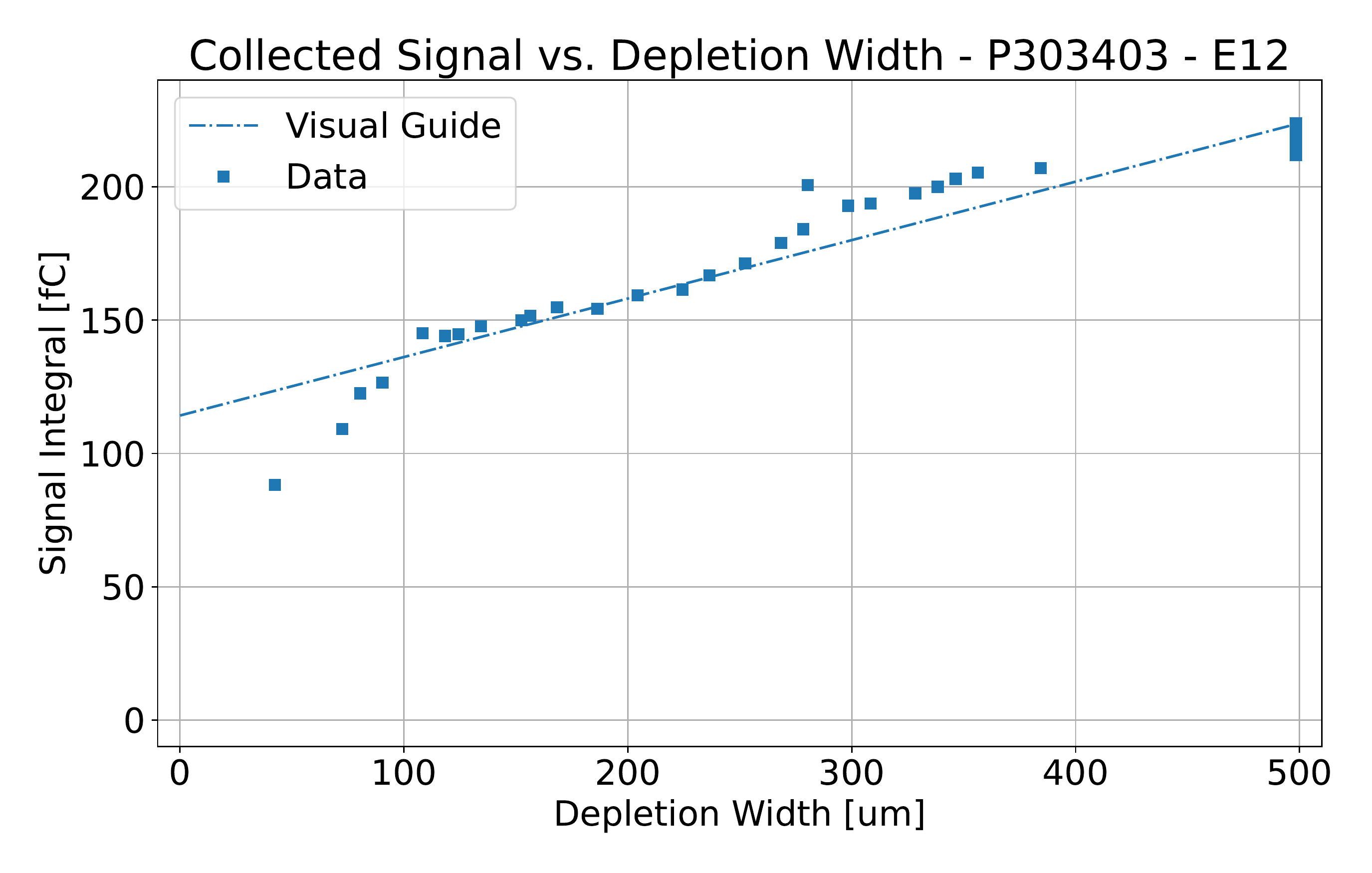}

  \caption{Maximum collected signal plotted against the extracted depletion width in the P-side. Each point represents a specific bias voltage.}
  \label{fig:depletion_vs_signal}
\end{figure}

In the following, we present some more details from the measurement of \textit{P303403-E12}.
From the data presented in figure~\ref{fig:bias_scan_plots} we can extract the signal collected within the depleted zone for each bias voltage.
Based on the prompt current data we estimate the depletion zone within the P-side at each bias voltage.
The resulting curve showing the signal collected in function of the depletion width is shown in figure~\ref{fig:depletion_vs_signal}.
We observe a proportional relationship between the depletion width and the maximum collected charge for this sample.

Figure~\ref{fig:timedomain_curves} shows a sample set of time domain TCT curves.
One observes a long exponential tail with a characteristic time $t \approx 25~\text{ns}$, which is present on all curves.
This tail shows very little variation with different bias voltages.
Reference measurements with a spaghetti diode supplied by Particulars~\cite{kramberger_charge_2013} do not show such an exponential tail.
This confirms that the tail is not due to the setup or due to the electronics front-end, but only is present with our bonded samples.

\section{Conclusions}
In this paper we presented the transient current technique setup at ETH Zürich, capable of doing edge TCT measurements using red and IR laser light.
In order to facilitate the sample handling we created a sandwich-type sample holder.
The special holder integrates shielding against noise pickup by creating a Faraday cage around the sample.
The previously unexplained decrease in collected signal at high bias voltages, which was observed in~\cite{wuthrich_depletion_2022}, is shown to be due to non-linear behaviour in the front-end electronics of the setups, which can be suppressed by operating the front-end amplifier at its maximum amplification.
The reviewed results strongly hint towards an asymmetric behaviour, indicating that only the P-side of our P-N samples is being depleted under reverse bias.

We investigated this asymmetric behaviour via IR edge TCT scans of six samples at various bias voltages.
The edge TCT measurements clearly confirm that no depletion of the N-side occurs.
This points to the interpretation that the electrons necessary to deplete the P-side are coming from the defects located at the bonding interface.
In the time domain TCT curves a long exponential tail is observed with a decay time of $\tau \approx 25~\text{ns}$.

\acknowledgments
We acknowledge Sebastian Pape from CERN / TU~Dortmund University for fruitful discussions about TCT analysis procedures and Matias Senger from UZH / PSI for inspirations to improve the TCT setup~\cite{senger_e_2022}.
For data processing and plotting we rely on the Python programming language using the Numpy~\cite{harris_array_2020}, SciPY~\cite{virtanen_scipy_2020} and Matplotlib~\cite{hunter_matplotlib_2007} packages.
Computer numerical design was carried out with FreeCad~\cite{riegel_freecad_2021} and KiCad~\cite{kicad_developers_team_kicad_2020} and vector graphics are created with Inkscape~\cite{inkscape_project_inkscape_2019}.

\bibliographystyle{JHEP}
\bibliography{BondedDiodeEdgeTCT_JWuethrich}

\end{document}